\documentclass[12pt]{article}

\begin{document}
\renewcommand{\thefootnote}{\fnsymbol{footnote}}

\begin{titlepage}
\hfill\parbox{4cm}{ KIAS-P00079 \\ hep-th/0012247 \\ December 2000}
\vspace{15mm}
\baselineskip 8mm
\begin{center}
{\Large \bf
Supersymmetry of Green-Schwarz Superstring \\
and Matrix String Theory
}
\end{center}

\baselineskip 6mm
\vspace{5mm}
\begin{center}
Seungjoon Hyun\footnote{\tt hyun@kias.re.kr} and
Hyeonjoon Shin\footnote{\tt hshin@kias.re.kr}
\\[5mm]
{\it School of Physics, Korea Institute for Advanced
Study, Seoul 130-012, Korea}
\end{center}

\thispagestyle{empty}


\vfill

\begin{center}
{\bf Abstract}
\end{center}
\noindent
We study the dynamics of Green-Schwarz superstring on the
gravitational wave background corresponding to the Matrix string
theory and the supersymmetry transformation rules of the superstring.
The dynamics is obtained in the light-cone formulation and is shown to
agree with that derived from the Matrix string theory. The
supersymmetry structure has the corrections due to the effect of the
background and is identified with that of the low energy one-loop
effective action of Matrix string theory in two superstring background
in the weak string coupling limit.
\vspace{20mm}
\end{titlepage}

\baselineskip 6.6mm
\renewcommand{\thefootnote}{\arabic{footnote}}
\setcounter{footnote}{0}
\renewcommand{\theequation}{\thesection .\arabic{equation}}
\newcommand{\what}{\widehat}
\newcommand{\wtilde}{\widetilde}


\section{Introduction}
\setcounter{equation}{0}

In the context of Matrix theory, the eleven dimensional M theory in
the infinite momentum frame \cite{ban43} or the discrete light-cone
quantized (DLCQ) M theory \cite{sus80} is described non-perturbatively
by the supersymmetric quantum mechanical system with 16
supersymmetries.  Upon toroidal compactification, the DLCQ M theory on
$p$ torus with $p \le 3$ has a description in terms of $p+1$
dimensional super Yang-Mills (SYM) theory on dual $p$ torus
\cite{sei9}.  In the dual supergravity description, they can be
described by M/Superstring theory on the appropriate supergravity
background.  For the DLCQ M theory, we thus have two discriptions; the
SYM and the supergravity theory.  In the supergravity side, the
process of having the DLCQ M theory and its compactified theory on $p$
torus leads us to the supergravity/superstring theories in certain
backgrounds as shown in Ref.~\cite{hyu26}.

For the uncompactified case, it has been shown that the leading order
dynamics from the low energy effective action of Matrix theory agrees
well with that of the classical supergravity (For a review, see for
example \cite{tay16} and references therein).  This remarkable
agreement between these two descriptions is basically due to the fact
that we have enough amount of supersymmetry, 16 supersymmetries
\cite{hyu22}.  Though it is true that the supersymmetry alone does not
give all possible dynamics of the theory, this leads us to consider an
issue about to what extent the supersymmetry restricts the dynamics.
Concerning this issue, one of the present authors \cite{hyu119} has
considered the Matrix theory in the supergravity side and obtained the
supersymmetry transformations rules for the eleven dimensional
supergraviton on the lifted D0-brane background, which correspond to
those of the low energy one-loop effective action of Matrix theory for
two supergraviton background.

In this paper, we are concerned about the Matrix string theory.  We
study the dynamics of superstring, which is the parton of the theory,
and the supersymmetry in the supergravity side.  The resulting
dynamics will be compared with that from the Matrix string theory
\cite{dij30} which is the SYM description of DLCQ M theory on a
circle.

Since the background corresponding to the Matrix string theory is
curved as noted above, we need the superstring action in type IIA
supergravity background for our purpose.  The superstring action
should be of Green-Schwarz (GS) type because the Matrix string theory
is the free GS light-cone superstring at its conformal point.  The
desired action expanded up to quadratic order in terms of the
anticommuting coordinates has been reported in Ref.~\cite{cve202},
which will be presented in Sec.~\ref{string2}.  As consistency checks
of it which have not been done in Ref.~\cite{cve202}, we will show
that it is supersymmetric and invariant under the $\kappa$-symmetry
transformation.

The other sections are organized as follows: In Sec.~\ref{lcstr}, we
study the dynamics of superstring in the background corresponding to
the Matrix string theory and compare the results with those from
the Matrix string theory.  The light-cone gauge is natural for our
purpose, and the phase space approach of \cite{god109} is adopted for
the light-cone formulation of superstring. We would like to note that,
in a recent work \cite{met171}, the phase space approach also has been
well applied in the program of quantizing the GS superstring in AdS$_5
\times S^5$ \cite{met28}.  In Sec.~\ref{susylc}, we investigate the
supersymmetry transformations rules which have corrections due to the
effect of the background and the supersymmetry algebra. The
identification of the supersymmetry structure with that of the Matrix
string theory effective action for two superstring background is
discussed.  Finally, discussion follows in Sec.~\ref{dis}.

\section{Green-Schwarz Superstring Action}
\setcounter{equation}{0} \label{string2}

In this section, we review the ten dimensional type IIA GS
superstring action in the bosonic supergravity backgrounds
constructed in Ref.~\cite{cve202} with the aim of fixing our
notations and for the self-containedness, and investigate its
symmetries: supersymmetry and $\kappa$-symmetry.

We begin with the superstring action embedded in the ten dimensional
target superspace.\footnote{The index notations adopted here are as
follows: $M,N,...$ are used for the target superspace indices while
$A,B,...$ for tangent superspace. As usual, a superspace index is the
composition of two types of indices such as $M=(\mu,\alpha)$ and
$A=(r,a)$. $\mu,\nu,...~(r,s,...)$ are the ten dimensional target
(tangent) space-time indices taking values in $0,1,...,9$.
$\alpha,\beta,...~(a,b,...)$ are the ten dimensional (tangent) spinor
indices with values in $1,2,...,32$. $m,n,...$ are the worldsheet
vector indices with values $\tau$ and $\sigma$. The convention for
the worldsheet antisymmetric tensor is taken to be
$\epsilon^{\tau\sigma}=1$.}
\begin{equation}
S =  \frac{1}{2\pi \alpha'} \int d^2 \sigma
  \left( \, -\frac{1}{2} \sqrt{-\gamma} \gamma^{mn}
            \Pi^r_m \Pi^s_n \eta_{rs}
        +\frac{1}{2!} \epsilon^{mn} \Pi^A_m \Pi^B_n B_{BA} \,
  \right)~,
\label{str}
\end{equation}
where $\gamma_{mn}$ is the string worldsheet metric, $\eta_{rs}$ is
the ten dimensional flat target space-time metric. $\Pi^A_m$ is the
pullback of the super zehnbein $E_M^{~A}$ onto the string worldsheet,
with the expression
\begin{equation}
\Pi^A_m =\partial_m Z^M E_M^{~A}~, \label{p10}
\end{equation}
and $B_{AB}$ is the second-rank antisymmetric tensor superfield.
$Z^M$ are the supercoordinates of the target superspace and are
denoted by $Z^M=(X^\mu,\theta^\alpha)$, where $\theta^\alpha$ is the
anticommuting coordinates, the 32 component Majorana spinor.
(Although, using the $\Gamma^{11}$ matrix, the ten dimensional
chirality operator, we can split $\theta$ into two Majorana-Weyl
spinors with opposite chiralities with 16 independent components, we
will keep $\theta$ to be Majorana for a while.)

The action (\ref{str}) is the one expressed in the context of
superfield formalism.  For practical applications, it is less useful
in its form and should be expanded in terms of anticommuting
coordinates $\theta$.  The expansion coefficients, the component
fields, are the functions of the ten dimensional type IIA supergravity
fields.  However, the component field expansion is a formidable task
basically because of many supergravity fields; 5 types even in the
bosonic sector only.  A rather easy way for obtaining the expansion is
given by the fact \cite{duf70} that the ten dimensional type IIA GS
superstring is related to the eleven dimensional supermembrane
\cite{ber330} through the double dimensional Kaluza-Klein reduction.
Using this fact, the authors of Ref.~\cite{cve202} have constructed
the type IIA GS superstring action in the bosonic type IIA
supergravity background starting from the supermembrane action
expanded up to the quadratic order in $\theta$ \cite{dew209}.  Though
the action is not a fully expanded one and has couplings only to the
bosonic backgrounds, it is enough and suitable for our purpose.

Before presenting the action, we give the component expansion of the
pullback of the super zehnbein up to $\theta^2$ order, which will be
used in the discussion of symmetries and in the later sections.
Following the double dimensional Kaluza-Klein reduction, we can
obtain, from the expansion of super elfbein \cite{dew209} with
vanishing fermionic backgrounds,\footnote{For the vanishing fermionic
backgrounds as in this paper, the order of $\theta$ increases by two
for all quantities having expansion in terms of $\theta$; if the
leading order term is of the even (odd) order in $\theta$, all the
higher order terms are of the even (odd) order in $\theta$.}  the
following expansion of the pullback of super zehnbein:\footnote{In our
convention, $\bar{\theta}=\theta^{\rm T} \Gamma^0$.}
\begin{equation}
\Pi^r_m = \partial_m X^\mu e_\mu^{~r}
            + i \bar{\theta} \Gamma^r \partial_m \theta
            + i \partial_m X^\mu
               (\bar{\theta} \Gamma^r \Omega_\mu \theta)
            + \frac{i}{2} \partial_m X^\mu e_\mu^{~r}
              (\bar{\theta} \Gamma^{11} \Omega_{11} \theta )
            + {\cal O} (\theta^4) ~,
            \label{p10c}
\end{equation}
where we have defined
\begin{eqnarray}
\Omega_\mu &=& \frac{1}{4} \omega_\mu^{~rs} \Gamma_{rs}
    +\frac{1}{6} \Gamma_{rs} e^{\nu [ r} e_\mu^{~s]} \partial_\nu \phi
    +\frac{1}{4} \Gamma^\nu \Gamma^{11} e^\phi F_{\mu \nu}
    + T_\mu^{~\: \nu \rho \sigma} \Gamma^{11} H_{\nu \rho \sigma}
                    \nonumber \\
& &    + T_\mu^{~\: \nu \rho \sigma \kappa}
        e^\phi F'_{\nu \rho \sigma \kappa}~,
                    \nonumber \\
\Omega_{11} &=&   -\frac{1}{3} \Gamma^\mu \Gamma^{11} \partial_\mu
\phi
     -\frac{1}{8} \Gamma^{\mu \nu} e^\phi F_{\mu \nu}
      + \frac{1}{288}
         ( 8 \Gamma^{\mu \nu \rho} H_{\mu\nu\rho}
          +\Gamma_{11} \Gamma^{\mu\nu\rho\sigma}
             e^\phi F'_{\mu\nu\rho\sigma} )~.
\label{om}
\end{eqnarray}
Here $e_\mu^{~r}$ is the zehnbein, $\omega_\mu^{~rs}$ the spin
connection, and $\phi$ the dilaton.  $H_{\mu\nu\rho}$ is the field
strength of the NS-NS antisymmetric second-rank tensor $B_{\mu\nu}$
field.  $F_{\mu\nu}$ and $F_{\mu\nu\rho\sigma}$ are the field
strengths of the R-R fields $A_\mu$ and $A_{\mu\nu\rho}$ related to
D-branes.  The field strength $F'_{rstu}$ is the modified gauge
invariant 4-form field strength defined by
\[
F'_{\mu\nu\rho\sigma} = F_{\mu\nu\rho\sigma} + 4 A_{[\mu}
H_{\nu\rho\sigma]} ~.
\]
$\Gamma^r$ is the ten dimensional Dirac gamma matrices and the
following tensor structures have been defined in Eq.~(\ref{om}):
\begin{eqnarray*}
T_\mu^{~\: \nu \rho \sigma \lambda} &\equiv& \frac{1}{288}
(\Gamma_\mu^{~\: \nu \rho \sigma \kappa}
        - 8 \delta_\mu^{[\nu} \Gamma^{\rho \sigma \kappa ]} ) ~,
                    \\
T_\mu^{~\: \nu \rho \sigma} &\equiv& \frac{1}{72} (\Gamma_\mu^{~\:
\nu \rho \sigma}
        - 6 \delta_\mu^{[\nu} \Gamma^{\rho \sigma ]} ) ~,
\end{eqnarray*}
where $\Gamma^{\mu\nu...}$ is the totally antisymmetric products of
the Dirac gamma matrices.

In writing the IIA GS superstring action constructed in
\cite{cve202}, we split Majorana spinor $\theta$ into two
Majorana-Weyl spinors with opposite chiralities, $\theta^1$ and
$\theta^2$, by using the gamma matrix $\Gamma^{11}$.  We assign
positive chirality to $\theta^1$ and negative chirality to
$\theta^2$;
\begin{equation}
\Gamma^{11} \theta^1 =\theta^1~,~~~ \Gamma^{11} \theta^2 =-\theta^2~.
\end{equation}
Then the type IIA  GS superstring action $S$ in non-trivial
bosonic supergravity backgrounds, expanded up to the quadratic order
in the anticommuting coordinates, is
\begin{equation}
S = S_{\rm kin} + S_{\rm WZ} ~, \label{2ags}
\end{equation}
where $S_{\rm kin}$ and $S_{\rm WZ}$ are the kinetic and the
Wess-Zumino (WZ) part, respectively.  Their expressions are as
follows:
\begin{eqnarray}
S_{\rm kin} &=&
 -\frac{1}{4 \pi \alpha'} \int d^2 \sigma \sqrt{-\gamma} \gamma^{mn}
 ( \partial_m X^\mu + i \bar{\theta}^I \Gamma^\mu \partial_m\theta^I )
 ( \partial_n X^\nu + i \bar{\theta}^J \Gamma^\nu \partial_n\theta^J )
   G_{\mu\nu}
                                        \nonumber \\
& &
 -\frac{i}{4 \pi \alpha'} \int d^2 \sigma \sqrt{-\gamma} \gamma^{mn}
 \partial_m X^\mu \partial_n X^\nu
                    \nonumber \\
& & \hspace{13mm} \times
 \bigg[ \:
     \frac{1}{2} (\bar{\theta}^I \Gamma_{\mu rs} \theta^I)
                   \omega_\nu^{~rs}
    -\frac{1}{4} s^{IJ} ( \bar{\theta}^I
                   \Gamma_\mu^{~\, \rho \sigma}
                   \theta^J) H_{\nu \rho \sigma}
                    \nonumber \\
& & \hspace{19mm}
    +\frac{1}{4}  (\bar{\theta}^1 \Gamma^{\rho\sigma} \theta^2 )
              e^\phi F_{\rho \sigma} G_{\mu\nu}
    - (\bar{\theta}^1 \Gamma_\mu^{~\, \rho} \theta^2)
              e^\phi F_{\nu \rho}
                                    \nonumber \\
& & \hspace{19mm}
    + \frac{1}{48}  ( \bar{\theta}^1
             \Gamma^{\rho \sigma \kappa \lambda} \theta^2)
             e^\phi F'_{\rho\sigma\kappa\lambda} G_{\mu\nu}
    -\frac{1}{6}  (\bar{\theta}^1
             \Gamma_\mu^{~\,\rho\sigma\kappa} \theta^2)
             e^\phi F'_{\nu \rho\sigma\kappa}
    +{\cal O}(\theta^4) \,
 \bigg] ~,
                    \nonumber \\
& &             \label{sk}
\end{eqnarray}
where $G_{\mu\nu}$ is the target space-time metric, $I,J= 1,2$ and
$s^{IJ}$ is defined as
\[
s^{11} = - s^{22} = 1~,~~~~s^{12}=s^{21}=0 ~,
\]
and
\begin{eqnarray}
S_{\rm WZ} &=& - \frac{i}{2 \pi \alpha'} \int d^2 \sigma \,
       \epsilon^{mn} s^{IJ}
       \left(
              \partial_m X^\mu
             +\frac{i}{2} \bar{\theta}^K \Gamma^\mu \partial_m
                          \theta^K \right)
        ( \bar{\theta}^I \Gamma_\mu \partial_n \theta^J )
                                        \nonumber \\
& &  - \frac{1}{4\pi \alpha'} \int d^2 \sigma \, \epsilon^{mn}
   \partial_m X^\mu \partial_n X^\nu
                    \nonumber \\
& & \hspace{13mm} \times
 \bigg[ \:
    B_{\mu\nu}
    +\frac{i}{2} s^{IJ} ( \bar{\theta}^I \Gamma_{\mu rs} \theta^J)
                 \omega_{\nu}^{~rs}
    -\frac{i}{4}(\bar{\theta}^I \Gamma_{\mu}^{~~\rho \sigma}
        \theta^I) H_{\nu  \rho \sigma}
                    \nonumber \\
& & \hspace{19mm}
    + \frac{i}{2} (\bar{\theta}^1 \theta^2) e^\phi F_{\mu\nu}
    + \frac{i}{4} ( \bar{\theta}^1 \Gamma_{\mu\nu}^{~~~\rho \sigma}
                  \theta^2 ) e^\phi F_{\rho \sigma}
                                                  \nonumber \\
& & \hspace{19mm}
    +\frac{i}{4} (\bar{\theta}^1 \Gamma^{\rho \sigma} \theta^2)
        e^\phi F'_{\mu \nu \rho \sigma}
    +\frac{i}{48} (\bar{\theta}^1
        \Gamma_{\mu \nu}^{~~~\rho \sigma \kappa \lambda} \theta^2)
        e^\phi F'_{\rho \sigma \kappa \lambda}
    + {\cal O} (\theta^4) \,
 \bigg] ~.  \label{swz}
\end{eqnarray}

There are several notable features in these actions. First of all, as
expected, we have explicit $e^\phi$ coupling in the terms linear in
R-R fields, $A_{\mu}$ and $A_{\mu\nu\rho}$, which has been first
suggested by Tseytlin \cite{tse109}.  Furthermore, as fundamental strings
are neutral under R-R fields, it is natural to couple with R-R fields,
if any, via their field strength. In particular, considering these
actions as describing the interactions of the fundamental string with
the background supergravity fields, these couplings have a natural
interpretation as a spin-orbit like coupling with background R-R
fields and imply that the fundamental string has dipole interactions
with R-R fields.  Indeed the interactions with the Lorentz
spin-connection give the genuine spin-orbit coupling between the
string and the gravitational backgrounds and has been extensively
studied with regard to the Matrix theory
\cite{hyu22,bal37,tay239,hyu105,kra199}.

Note also that, at the linearized level in the supergravity fields
$B_{\mu\nu}$, $A_\mu$, $A_{\mu\nu\rho}$, $g_{\mu\nu}$ and $\varphi$,
where $G_{\mu\nu}=\eta_{\mu\nu}+ g_{\mu\nu}$ and $\phi=\phi_\infty
+\varphi$, the action can be thought as the sum of free string action
and vertex operators for emission of bosonic supergravity
fields.\footnote{ Recent construction of vertex operators in the GS
superstring theory is given in Refs.~\cite{gre155,das280}.}  In this
sense it is very natural that the vertex operators for R-R fields
starts with $\theta^2$ order as there are target space-time
supersymmetries connecting NS sector and R sector. Schematically,
under space-time supersymmetry transformations in their leading order
in $\theta$, the vertex operators $V_{\rm NS-NS}$ and $V_{\rm R-R}$
corresponding to the fields in NS-NS and R-R sector, respectively,
should satisfy
\[
\delta_{B}V_{\rm NS-NS}+\delta_{F}V_{\rm R-R}=0~,
\]
where subscripts $B$ and $F$ denote supersymmetry transformation of
bosonic coordinate $X^\mu$ and fermionic coordinate $\theta$,
respectively. In the following subsection, we will see this is indeed
the case.

\subsection{Local supersymmetry}
\label{susy}

In this and the next subsection, we investigate the invariance of the
superstring action $S$, Eq.~(\ref{2ags}), under the supersymmetry and
$\kappa$-symmetry transformation in order to check whether or not the
action was correctly expanded.  We note that, for the simplicity of
presentation, we will not split Majorana spinors $(\theta, \eta,
\kappa)$ into Majorana-Weyl spinors.

The local supersymmetry, super diffeomorphism, is the local change of
supercoordinates $(\delta_\eta X^\mu, \delta_\eta
\theta^\alpha)$. Since its parameters are superfields, the
transformations of supercoordinates have component expansion in terms
of $\theta$.  In the eleven dimensional case, the expansions of the
parameters have been given in \cite{dew209}.  With the vanishing
fermionic backgrounds, the Kaluza-Klein reduction of them to ten
dimensions leads to
\begin{eqnarray}
\delta_\eta X^\mu &=& i\bar{\theta} \Gamma^\mu \eta
        +{\mathcal O}(\theta^3)~,     \nonumber \\
\delta_\eta \theta &=& \eta
        +{\mathcal O} (\theta^2) ~. \label{susyt}
\end{eqnarray}
The terms on the right-hand sides are enough for the transformation of
the superstring action expanded up to $\theta^2$ order, since higher
order corrections in the transfomation rules require terms of higher
order than $\theta^2$ order in the action.  Furthermore, the
supersymmetry variation of the action is valid up to the linear order
in $\theta$, because the transformation $\delta_\eta X^\mu$ acting on
the terms of $\theta^2$ order requires the $\theta^3$ order terms in
the action.

The superstring action contains background fields as well as
supercoordinates.  Thus the action is in fact not invariant with only
the above transformations, Eq.~(\ref{susyt}).  As noted by the authors
of Ref.~\cite{dew209}, the invariance of the action means that the super
diffeomorphsim induces the supersymmetry transformations of the
background fields. In other words, the action is supersymmetric if its
variation under the super diffeomorphism vanishes modulo supersymmetry
transformations of the background fields. This fact requires us to
have the supersymmetry transformation rules for the background
fields. What we need are those of fermion background fields, because
the transformation rules of bosonic fields lead to fermionic fields
which are turned off in this paper.  Fermion fields in ten dimensional
supergravity are the gravitino, $\psi_\mu$, and the dilatino,
$\lambda$, which are Majorana fermions and split into two
Majorana-Weyl fermions with opposite chiralities, $\psi_\mu^{1,2}$ and
$\lambda^{1,2}$, respectively.  We note that the transformation rules
of them are to be written in the string frame, since the object
affected by the backgrounds is the string.  Under the supersymmetry
variation, the gravitino transforms as
\begin{eqnarray}
\delta_\eta \psi_\mu
 &=&
   D_\mu ( \omega ) \eta
   -\frac{1}{8} \Gamma_r \Gamma_s \eta
        e_\mu^{~r} e^{\nu s} \partial_\nu \phi
                                                \nonumber \\
 & &
  -\frac{1}{64} e_\mu^{~r} ( \Gamma_{rst}- 14 \eta_{rs} \Gamma_t)
     \Gamma^{11} \eta e^\phi F^{st}
                                                \nonumber \\
 & & +\frac{1}{96} e_\mu^{~r}
     (\Gamma_r^{~stu}-9 \delta_r^s \Gamma^{tu}) \Gamma^{11} \eta
      H_{stu}
                                                \nonumber \\
 & &
     +\frac{1}{768} e_\mu^{~r}
     ( 3\Gamma_r^{~stuv}-20 \delta_r^s \Gamma^{tuv} ) \eta
     e^\phi F'_{stuv} ~,                        \label{dpsi}
\end{eqnarray}
where the Lorentz covariant derivative, $D_\mu (\omega)$, is given by
$D_\mu (\omega)= \partial_\mu + \frac{1}{4} \omega_\mu^{~rs}
\Gamma_{rs}$. As for the dilatino field, the supersymmetry
transformation rule is
\begin{eqnarray}
\delta_\eta \lambda
 &=& -\frac{1}{2 \sqrt{2} } \Gamma^r \Gamma^{11} \eta
            e_r^{~\nu} \partial_\nu \phi
    -\frac{3}{16 \sqrt{2} } \Gamma^{rs} \eta e^\phi F_{rs}
                                               \nonumber \\
 & & + \frac{1}{24 \sqrt{2}} \Gamma^{rst} \eta H_{rst}
    + \frac{1}{192 \sqrt{2} } \Gamma^{rstu} \Gamma^{11}
            \eta  e^\phi F'_{rstu} ~.
                                               \label{dlam}
\end{eqnarray}
These transformation rules are those with the vanishing fermion
backgrounds.  In the study of supergravity, the transformation rules
are usually written in the Einstein frame \cite{huq112}, which can be
obtained from the above transformation rules by a suitable rescaling
of fields and supersymmetry parameter. The resulting transformation
rules are the same as above except for the absence of the term
involving the derivative of dilaton in Eq.~(\ref{dpsi}) and some
change in powers of dilaton factor.

For the supersymmetry variation of the kinetic term, Eq.~(\ref{sk}),
it is enough to consider the variation of the pullback of
super-zehnbein under the super diffeomorphism, Eq.~(\ref{susyt}). A
straightforward calculation shows that
\begin{equation}
\delta_\eta \Pi^r_m =  i \partial_m X^\mu
      \left(
        2 \bar{\theta} \Gamma^r \delta_\eta \psi_\mu
        + \frac{1}{\sqrt{2}} e_\mu^{~r}
                \bar{\theta} \Gamma^{11} \delta_\eta \lambda
      \right)
     - \Lambda^r_{~s} \Pi^s_m + {\cal O} (\theta^3)~,
\label{dpi}
\end{equation}
where $\Lambda^r_{~s}$ is the local Lorentz transformation parameter
associated with the super diffeomorphism and is precisely given by
\begin{eqnarray}
\Lambda^{rs} &=& i(\bar{\theta} \Gamma^\mu \eta) \,
\omega_{\mu}^{~rs}
    -\frac{1}{3} i \bar{\theta}
                 (\Gamma^{rs} \Gamma^t-\Gamma^{rst}) \eta \,
             e_t^{~\mu} \partial_\mu \phi
                                        \nonumber \\
& &  +\frac{1}{3 \sqrt{2}}i \bar{\theta} \Gamma^{rs} \Gamma_{11}
                  \delta_\eta  \lambda
     -\frac{1}{2}i (\bar{\theta} \Gamma_{11} \eta )
               e^\phi F^{rs}
                                        \nonumber \\
& &  +\frac{1}{36} \,
      i \bar{\theta} \,
          (  \Gamma^{rstuv} \Gamma^{11} H_{tuv}
            +12 \Gamma_t \Gamma_{11} H^{rst}
          ) \, \eta
                                        \nonumber \\
& &  +\frac{1}{144} \,
     i \bar{\theta} \,
          ( \Gamma^{rstuvw} e^\phi F'_{tuvw}
            +24 \Gamma_{tu} e^\phi {F'}^{rstu}
          ) \, \eta + {\cal O} (\theta^3) ~.
 \label{lorentz}
\end{eqnarray}
Though it has a little bit complicated expression, $\Lambda^r_{~s}$
does not enter into the variation of the kinetic part, since two super
zehnbeins enter into the kinetic part symmetrically. We now see the
supersymmetry transformations of fermion backgrounds, $\delta_\eta
\psi_\mu$ and $\delta_\eta \lambda$, which have been identified with
the right-hand sides of Eqs.~(\ref{dpsi}) and (\ref{dlam}) at the
final stage of deriving Eqs.~(\ref{dpi}) and (\ref{lorentz}). However,
this does not conclude that the component expansion of the pullback of
super zehnbein or the kinetic part has been obtained correctly, since
the expansion has been given with the vanishing fermion
backgrounds. Thus, we turn on the fermion backgrounds temporarily and
investigate how they appear in the super zehnbein.  What we should be
concerned about are the linear order terms in $\theta$, which can be
seen by looking at the super elfbein expanded in
Ref.~\cite{dew209}. The corresponding term in the expansion of the
super elfbein is $2 i \bar{\theta} \Gamma^{\hat{r}}
\hat{\psi}_{\hat{\mu}}$, where $\hat{\psi}$ is the eleven dimensional
gravitino, and $\hat{r}$ and $\hat{\mu}$ are the eleven dimensional
flat and curved indices respectively.  The Kaluza-Klein dimensional
reduction of it can be done in the usual manner,\footnote{Through the
Kaluza-Klein reduction, the eleven dimensional gravitino is related to
the ten dimensional gravitino $\psi_\mu$, the dilatino $\lambda$, and
the dilaton $\phi$ as follows: $\hat{\psi}_\mu = e^{-\phi/6} \left(
\psi_\mu - \frac{\sqrt{2}}{12} \Gamma_\mu \Gamma^{11} \lambda
\right)$, $\hat{\psi}_{11} = \frac{2 \sqrt{2}}{3} e^{5\phi/6}
\lambda$.  Here the ten dimensional quantities are those in the string
frame.} and leads to
\begin{equation}
2 i \bar{\theta} \Gamma^r \psi_\mu +
\frac{i}{\sqrt{2}} \bar{\theta} \Gamma^{11} \lambda e_\mu^{~r} -
\frac{i}{3 \sqrt{2}} \bar{\theta} \Gamma^r_{~\mu} \Gamma^{11}
\lambda ~.
\end{equation}
We see that the supersymmetry variation of the fermion backgrounds in
this result exactly matches with the terms appearing in
Eqs.~(\ref{dpi}) and (\ref{lorentz}).  This concludes that the kinetic
term of the superstring action has the correct component expansion and
is consistent with the local supersymmetry.

We now turn to the supersymmetry variation of the WZ term,
Eq.~(\ref{swz}).  After the same type of calculations with those for
the kinetic term, it is given by
\begin{eqnarray}
\delta_\eta \left(
          \frac{1}{2!} \epsilon^{mn} \Pi^A_m \Pi^B_n B_{BA}
       \right)
 &=& - i \epsilon^{mn} \partial_m X^\mu \partial_n X^\nu
    \left(
        2  \bar{\theta} \Gamma^{11} \Gamma_\nu
           \delta_\eta \psi_\mu
     + \frac{1}{\sqrt{2}}
        \bar{\theta} \Gamma_{\nu \mu} \delta_\eta \lambda
    \right)                                 \nonumber \\
 & & + \partial_m S^m + {\cal O} (\theta^3) ~,
   \label{wzt}
\end{eqnarray}
where $\partial_m S^m$ is a surface term, which can be ignored since
we are concerned about the closed string.  This is the desired
result. However, as in the case of the kinetic term, we need to know
the terms of linear order in $\theta$ containing the fermion
backgrounds in order to check the supersymmetry of the WZ term. The
eleven dimensional term relevant to them comes from the component
expansion of the third-rank tensor superfield, and is given by $- 6i
\bar{\theta} \Gamma_{ [ \tilde{11} \mu} \hat{\psi}_{\nu ]}$
\cite{dew209}. Through the dimensional reduction, it reduces to
\begin{equation}
- 4i \bar{\theta} \Gamma_{11} \Gamma_{[ \mu}
\psi_{\nu ]} - \sqrt{2} i \bar{\theta} \Gamma_{\mu \nu} \lambda~.
\end{equation}
Obviously, the supersymmetry variation of this exactly matches with
the terms of Eq.~(\ref{wzt}). This tells us that the WZ term of the
superstring action as well as the kinetic term has the correct
component expansion and is consistent with the local supersymmetry.

We have seen that each term of the superstring action transforms
properly under the supersymmetry transformation, and has the
consistent and correct component expansion.

\subsection{$\kappa$-symmetry}

As another consistency check, we now consider the invariance of
the superstring action, Eq.~(\ref{2ags}), under the $\kappa$
transformation. The investigation of the $\kappa$ symmetry will give
us the confirmation on the correctness of relative coefficient
between the kinetic and the WZ part.

For the $\kappa$ symmetry, it is convenient to work with the
superstring action with Nambu-Goto type rather than the Polyakov type
action, Eq.~(\ref{str}), which avoids the complexity due to the
variation of the worldsheet metric $\gamma^{mn}$.  The Nambu-Goto type
action is obtained by solving the classical equation of motion for
$\gamma^{mn}$ and putting the result back into the action
Eq.~(\ref{str}), and is as follows:
\begin{equation}
S_{\rm NG} =  \frac{1}{2\pi \alpha'} \int d^2 \sigma
  \left( - \sqrt{-g}
         +\frac{1}{2!} \epsilon^{ij} \Pi^A_i \Pi^B_j B_{BA}
  \right) ~,
\label{ngstr}
\end{equation}
where $g$ is the determinant of the induced metric $g_{mn}$ given by
\begin{equation}
g_{mn} =   \Pi^r_m \Pi^s_n \eta_{rs}~.
\end{equation}

 We notice that the $\kappa$ transformation rules have also a
component expansion in terms of $\theta$.  As in the case of
supersymmetry, however, only the leading order terms in the expansion
are needed for showing that the superstring action constructed up to
$\theta^2$ order is $\kappa$ symmetric:
\begin{eqnarray}
\delta_\kappa X^\mu &=& i \bar{\kappa}_+ \Gamma^\mu \theta + {\cal O}
(\theta^3) ~,
                            \nonumber \\
\delta_\kappa \theta^a &=& \kappa^a_+ + {\cal O} (\theta^2) ~.
 \label{kappat}
\end{eqnarray}
where we have defined
\begin{equation}
\kappa_+ = ( 1+\Gamma \Gamma^{11}) \kappa ~. \label{kp}
\end{equation}
The matrix $\Gamma$ is given by
\begin{equation}
\Gamma = \frac{1}{2 \sqrt{-g}} \epsilon^{mn} \Pi_m^r \Pi_n^s
         \Gamma_r \Gamma_s
       = \frac{1}{2 \sqrt{-g}} \epsilon^{mn}
         \Gamma_m \Gamma_n~,
         \label{gam}
\end{equation}
where $\Gamma_m$ are the pullback onto the worldvolume of the
space-time Dirac gamma matrices:
\begin{equation}
\Gamma_m = \Pi_m^r \Gamma_r ~.
\end{equation}
$\Gamma$ has the properties as projection operator and anticommutes
with the pulled back gamma matrices $\Gamma_m$:
\[
\Gamma^2 = 1~,~~{\rm Tr} \Gamma = 0~,
\]
\begin{equation}
\Gamma \Gamma^m = -\Gamma^m \Gamma
  = -\frac{1}{\sqrt{-g}} \epsilon^{mn} \Gamma_n~.
\label{ggi}
\end{equation}

Under the $\kappa$ transformations, Eq.~(\ref{kappat}), the pullback
of the super zehnbein transforms as
\begin{equation}
\delta_\kappa \Pi_i^r
 =  2i \bar{\kappa}_+ \Gamma^r \partial_i \theta
     + i \Pi_i^s ( 2 \bar{\kappa}_+ \Gamma^r \tilde{\Omega}_s \theta
                +\bar{\kappa}_+ \Gamma^{11} \tilde{\Omega}_{11} \theta
                 \delta^r_s )
     + M^r_{~s} \Pi_i^s
     + {\cal O} (\theta^3) ~,
\end{equation}
where $\tilde{\Omega}_s~(\tilde{\Omega}_{11})$ is
$\Omega_s~(\Omega_{11})$ in Eq.~(\ref{om}) without terms involving the
derivatives of dilaton $\phi$. $M^r_{~s}$ is an antisymmetric matrix
just like the Lorentz transformation parameter $\Lambda^r_{~s}$ in
Eq.~(\ref{dpi}), whose detailed from is not necessary because it does
not give any contribution to the variation of the superstring
action. The $\kappa$ transformation of the kinetic term of the
Nambu-Goto type is then
\begin{eqnarray}
\delta_\kappa S_{\rm kin} &=& - \frac{i}{2 \pi \alpha'} \int d^2
\sigma \sqrt{-g} g^{mn}
  \Big[ \,
       2 \Pi_m^r \bar{\kappa}_+ \Gamma_r \partial_n \theta
                    \nonumber \\
& &  + \Pi_m^r \Pi_n^s
        ( 2 \bar{\kappa}_+ \Gamma_r \tilde{\Omega}_s \theta
         + \eta_{rs}
           \bar{\kappa}_+ \Gamma^{11} \tilde{\Omega}_{11} \theta )
      + {\cal O} (\theta^3)
  \Big]  ~.
  \label{nskt}
\end{eqnarray}
For the WZ term, we have the following $\kappa$ transformation.
\begin{eqnarray}
\delta_\kappa S_{\rm WZ} &=& \frac{i}{2\pi \alpha'} \int d^2 \sigma
\epsilon^{mn}
  \Big[ -2 \Pi_m^r \bar{\kappa}_+ \Gamma_r \Gamma^{11}
                    \partial_n \theta
                    \nonumber \\
& & + \Pi_m^r \Pi_n^s
    ( 2\bar{\kappa}_+
      \Gamma^{11}\Gamma_r \tilde{\Omega}_s \theta
     +\bar{\kappa}_+ \Gamma_{rs} \tilde{\Omega}_{11} \theta )
    +{\cal O} (\theta^3)
  \Big] ~.
  \label{nswzt}
\end{eqnarray}
With these transformations and by using some properties of the matrix
$\Gamma$, Eq.~(\ref{ggi}), we can now show that the Nambu-Goto type
string action, Eq.~(\ref{ngstr}), is $\kappa$ symmetric up to the
quadratic order in $\theta$:
\begin{eqnarray}
\delta_\kappa S_{\rm NG} &=& - \frac{i}{ \pi \alpha'} \int d^2 \sigma
\sqrt{ - g}
  \bar{\kappa}_+ (1-\Gamma\Gamma^{11})
  ( \Gamma^m \partial_m +\Gamma^m \tilde{\Omega}_m +
    \Gamma^{11} \tilde{\Omega}_{11}) \theta
  + {\cal O} (\theta^3)
                    \nonumber \\
&=& 0 + {\cal O} (\theta^3) ~,
\end{eqnarray}
where $\bar{\kappa}_+ = \bar{\kappa}(1+\Gamma\Gamma^{11})$ and
$(1+\Gamma\Gamma^{11}) (1-\Gamma\Gamma^{11}) = 0$ have been used.

\section{GS Superstring on the Gravitational Wave Background} 
\setcounter{equation}{0}
\label{lcstr}

In this section, by using the superstring action Eq.~(\ref{2ags})
presented in the previous section, we study the dynamics of
superstring on the gravitational wave background corresponding to the
Matrix string theory. From the Matrix string theory side calculations,
though the dynamics related to the fermion bilinears has not been
completely determined, some results to be compared with ours have been
reported \cite{sch145}.  As will be shown in this section, they agree
with the bosonic part of our result.

The ten dimensional supergravity background corresponding to the
Matrix string theory \cite{hyu26} is the geometry obtained after the
following procedure; one begins with the D0-brane geometry, follows
the prescription of Seiberg and Sen \cite{sei9} and taking the
TST-duality chain the same as that used for obtaining the Matrix
string theory \cite{dij30}.  In the resulting background, only the
metric is non-trivial.  The dilaton is just constant and all other
supergravity fields are simply zero.  If we introduce the light-cone
coordinates,
\begin{equation}
x^\pm = x^9 \pm x^0~,
\end{equation}
then the background geometry is as follows:
\[
ds^2 = dx^+ dx^- + h (dx^-)^2 + (dx^i)^2 ~,
\]
\begin{equation}
e^\phi = g_s~,
 \label{back}
\end{equation}
where $i$ is the index for the eight dimensional flat transverse space
taking values in $1,...,8$ and $g_s$ the string coupling constant As
an important structure of this geometry, the light-like direction
$x^-$ is compactified with the radius $R$, leading to the Discrete
Light-Cone Quantization (DLCQ):
\begin{equation}
x^- \sim x^- + 2 \pi n R~,~~~n \in {\bf Z}~.
\label{lcc}
\end{equation}
$h$ is the harmonic function in the eight dimensional transverse
space spanned by $x^i$ and is given by, with $r=(x^i x^i)^{1/2}$,
\begin{equation}
h = \frac{4}{\pi} \frac{g_s^2 l_s^8}{R^2} \frac{N_s}{r^6}~,
 \label{har}
\end{equation}
where $l_s$ is the string scale and $N_s$ the light-cone momentum of
the background.  In addition to the above background geometry, what we
need to write down the superstring action is the spin connection for
the geometry, whose non-vanishing components are given by
\begin{eqnarray}
 & &\omega^0_{~i}~=~\omega^i_{~0}~=~\omega^9_{~i}~=-\omega^i_{~9}
 ~=~ \frac{1}{2} f^{-1/2} \partial_i h dx^- ~,
                    \nonumber \\
 & &\omega^0_{~9}~=~\omega^9_{~0}
 ~=~ \frac{1}{2} f^{-1} \partial_i h dx^i ~,
 \label{spin}
\end{eqnarray}
where we have defined $f = 1+h$.

As is well known and shown explicitly in the previous section, the GS
superstring action has worldsheet local fermionic $\kappa$-symmetry,
which indicates the doubling of the degrees of freedom described by
$\theta$.  Our study of the string dynamics begins with the
consideration of fixing the fermionic symmetry. Since our basic
concern is the supergravity side description of the DLCQ M theory,
the light-cone gauge fixing condition is very natural. In the flat
background, the light-cone gauge fixing condition for the
$\kappa$-symmetry enables us to have a greatly simplified action. As
we shall see, this is also the case in the background (\ref{back}).
The $\kappa$-symmetry fixing condition we choose is then
\begin{equation}
\Gamma^+ \theta^I = 0~.
\label{kfix}
\end{equation}
In order to solve this, the representation for the $SO(1,9)$ Dirac
gamma matrices is in order.  The representation we take in this paper
is as follows:
\[
 \Gamma^0= i\sigma^2 \otimes {\bf 1}_{16}~,~~~
 \Gamma^9= \sigma^3 \otimes {\bf 1}_{16}~,~~~
 \Gamma^i= \sigma^1 \otimes \gamma^i~,
\]
\begin{equation}
\Gamma^{\pm} = \Gamma^9 \pm \Gamma^0~,
\end{equation}
where $\sigma$'s are Pauli matrices, and ${\bf 1}_{16}$ the $16 \times
16$ unit matrix. $\gamma^i$ are the $16 \times 16$ symmetric real
gamma matrices satisfying the spin$(8)$ Clifford algebra $\{ \gamma^i,
\gamma^j \} = 2 \delta^{ij}$, which are actually reducible to the
${\bf 8_s}+{\bf 8_c}$ representation of spin$(8)$.  With this
representation, the $\kappa$-symmetry fixing condition (\ref{kfix})
implies that $\theta^I$ has the following form:
\begin{equation}
\theta^I = \left( \begin{array}{c} \psi^I \\ -\psi^I \end{array}
       \right) ~,
 \label{tp}
\end{equation}
where $\psi^I$ is the 16 component Majorana-Weyl spinor.

The gauge condition for the $\kappa$-symmetry now simplifies the
superstring action in the background geometry (\ref{back}) as
\begin{eqnarray}
S
 &=& - \frac{1}{2} \int d^2 \sigma \sqrt{-\gamma} \,
     \bigg[ \,
    \gamma^{mn} ( \partial_m X^+ + h \partial_m X^- )
            \partial_n X^-
    + \gamma^{mn} \partial_m X^i \partial_n X^i
                                    \nonumber \\
 & & +8i f^{-1/2} (\partial_m X^+ + h \partial_m X^-) P^{mn,IJ}
    ( \psi^I \partial_n \psi^J)
                                    \nonumber \\
 & & + 4i f^{-1/2}  \partial_i h
    \partial_m X^j \partial_n X^- P^{mn,IJ}
    (\psi^I \gamma^{ij} \psi^J )
    + {\mathcal O} (\psi^4) \,
     \bigg]~,
\label{kgs}
\end{eqnarray}
where Eq.~(\ref{tp}) has been used and $P^{mn,IJ}$ defined by
\begin{equation}
P^{mn,IJ} \equiv \frac{1}{2}
    \left(
       \gamma^{mn} \delta^{IJ}
    + \frac{\epsilon^{mn}}{\sqrt{-\gamma}} s^{IJ}
    \right)
    \label{proj}
\end{equation}
are the projection tensors which project a worldsheet vector into its
self-dual $(I=J=1)$ or anti-self-dual $(I=J=2)$ pieces.  For
notational convenience, we have set $2 \pi \alpha' = 1$ in the action
(\ref{kgs}).  We note the overall factor $f^{-1/2}$ in the fermion
bilinear terms can be made disappear in the action simply via the
rescaling
\begin{equation}
\psi^I \longrightarrow f^{1/4} \psi^I ~.  \label{scale1}
\end{equation}
This is possible due to the fact that the Majorana-Weyl spinors
$\psi^I$ satisfy $\psi^I \psi^I = 0$.  In what follows, this rescaling
will be understood.

Having fixed the $\kappa$-symmetry, the action (\ref{kgs}) still has
an additional local symmetry, the worldsheet diffeomorphism.  We fix
this symmetry by taking the light-cone gauge.  Since the
diffeomorphism is two parameter symmetry, two conditions are required.
We choose the first to be that the light-cone time $X^+$ is
proportional to the worldsheet time $\tau$.  In the DLCQ framework,
since a certain sector of the light-cone momentum, canonical momentum
of $X^-$, is considered, it is convenient to have constant light-cone
momentum independent on the worldvolume spatial coordinate $\sigma$.
Thus it is natural to choose the second condition as constant
light-cone momentum.  This type of light-cone gauge, the same type as
in the case of flat background, is the one taken in what is known as
the phase space approach of string quantization \cite{god109}. The
phase space means that we should formulate our system in its phase
space because one of the gauge fixing conditions we choose is imposed
on a canonical momentum.  In order to impose our gauge fixing
condition, we should rewrite the Lagrangian in the phase space, that
is, in the first order form.  It should be noted here that we will not
touch the fermionic part, since the canonical momentum of the
fermionic coordinate is constraint, which will be treated later.  We
first obtain the canonical momenta of bosonic coordinates from
\[
 P^+ = \frac{\partial {\cal L}}{\partial \dot{X}^-}~,~~~
 P^- = \frac{\partial {\cal L}}{\partial \dot{X}^+}~,~~~
 P^i = \frac{\partial {\cal L}}{\partial \dot{X}^i}~,
\]
where the dot means the derivative with respect to the worldsheet time
$\tau$. In what follows, the prime will be used for the derivative
with respect to the worldsheet spatial coordinate $\sigma$.  The
explicit expressions for the canonical momenta are then as follows:
\begin{eqnarray}
P^+ &=& - \frac{1}{2}\sqrt{-\gamma} \,
    [ \, \gamma^{\tau\tau} (\dot{X}^+ + 2h \dot{X}^-)
    + \gamma^{\tau\sigma} (X'^+ + 2h  X'^-)
    + 8i h P^{\tau m,IJ}
        (\psi^I \partial_m \psi^J )
                    \nonumber \\
    & &
    + 4i \partial_i h \partial_m X^j  P^{m\tau,IJ}
        (\psi^I \gamma^{ij} \psi^J ) \,] +{\cal O} (\psi^4)~,
                    \nonumber \\
P^- &=& - \frac{1}{2} \sqrt{-\gamma} \,
    [ \, \gamma^{\tau\tau} \dot{X}^- + \gamma^{\tau\sigma} X'^-
      + 8i P^{\tau m,IJ}
        (\psi^I \partial_m \psi^J ) \,] +{\cal O} (\psi^4)~,
                    \nonumber \\
P^i &=& - \sqrt{-\gamma} \,
    [ \, \gamma^{\tau\tau} \dot{X}^i +  \gamma^{\tau\sigma} X'^i
     - 2i \partial_j h \partial_m X^-  P^{\tau m,IJ}
        (\psi^I \gamma^{ij} \psi^J ) \,] +{\cal O} (\psi^4)~.
\end{eqnarray}
From the action (\ref{kgs}), the Lagrangian in the first order form is
then given by
\begin{eqnarray}
{\cal L} &=& \dot{X}^+ P^- + \dot{X}^- P^+ + \dot{X}^i P^i
   + 4i \left( (P^+ - hP^- ) \delta^{IJ}
      + \frac{1}{2} (X'^+ + h X'^-) s^{IJ} \right)
        ( \psi^I \dot{\psi}^J )
                                        \nonumber \\
 & & + \frac{2}{\gamma^{\tau\tau} \sqrt{-\gamma}} {\cal H}_{00}
     + \frac{\gamma^{\tau\sigma}}{\gamma^{\tau\tau}} {\cal H}_{01}~,
     \label{fol}
\end{eqnarray}
where
\begin{eqnarray}
{\cal H}_{00} &=& P^- ( P^+ - hP^-) + \frac{1}{4} (P^i)^2
    + \frac{1}{4} \left( \, X'^- (X'^+ + h X'^-) + (X'^i)^2 \,
                  \right)
                    \nonumber \\
& & + 2i \left( \, (P^+ - h P^- ) s^{IJ}
              +\frac{1}{2} (X'^+ + hX'^-) \delta^{IJ} \,
         \right) (\psi^I \psi'^J)
                    \nonumber \\
& & - i \partial_i h
    \left( \, P^- ( P^j \delta^{IJ} + X'^j s^{IJ} )
      - \frac{1}{2} X'^- ( P^j s^{IJ}+ X'^j \delta^{IJ} )
    \right) ( \psi^I \gamma^{ij} \psi^J )
                            \nonumber \\
 & &    + {\cal O} (\psi^4)~,
\end{eqnarray}
\begin{eqnarray}
{\cal H}_{01}
 &=& P^- X'^+ + P^+ X'^- + P^i X'^i
                    \nonumber \\
 & & + i
    \left( \, (P^+ - h P^- ) \delta^{IJ}
        + \frac{1}{2} (X'^+ + h X'^- ) s^{IJ}
    \, \right) ( \psi^I \psi'^J)
    + {\cal O} (\psi^4)~.
\end{eqnarray}
In Eq.~(\ref{fol}), the worldsheet metric is non-dynamical and its
combinations, $1/ ( \gamma^{\tau\tau}\sqrt{-\gamma})$ and
$\gamma^{\tau\sigma}/\gamma^{\tau\tau}$, act as Lagrange multiplier
fields leading to the constraints
\begin{equation}
{\cal H}_{00} \approx 0~,~~~{\cal H}_{01} \approx 0~,
\label{virasoro}
\end{equation}
which are just the Virasoro constraints.  As usual, these enable us
to determine $P^-$ and $X'^-$ in terms of the canonical pairs of the
transverse coordinates, $(X^i,P^i)$, and the fermionic coordinates
$\psi^I$.

We now fix the worldsheet diffeomorphsim by the light-cone gauge
alluded above, which is explicitly given by
\begin{equation}
X^+ = 2 \tau ~,~~~ P^+ = p^+ l^{-1} = {\rm const.} \label{dfix}
\end{equation}
where $l$ is the range of $\sigma$ integration which we set equal to
one ($l=1$). $p^+$ is the center of mass momentum in the $X^-$
direction and its value is quantized as $p^+ = N/R$ ($N$ is an
integer and means that we are in the $N$ sector of DLCQ.) since $X^-$
direction is compactified as in (\ref{lcc}). After imposing this
light-cone gauge and the constraints Eq.~(\ref{virasoro}) in a strong
sense, we are left with the reduced system containing only the
physical degrees of freedom; $(X^i,P^i)$, $\psi^I$ and the canonical
pair corresponding to the center of mass mode of $X^-$, $(x^-, p^+)$.

In the phase space, the dynamics is described by the canonical
Hamiltonian.  For our light-cone gauge fixed system, it is just the
light-cone Hamiltonian and is given by
\begin{equation}
H \, = \, \dot{x}^- p^+ + \int^1_0 d \sigma \, ( \,
         \dot{X}^i P^i + \dot{\psi}^I P^I - {\cal L}
    \,)
  \, = \, - 2 \int^1_0 d \sigma \, P^-~,
\end{equation}
where $P^I$ is the canonical momentum of $\psi^I$ given by
\begin{equation}
P^I =  \frac{\partial {\cal L}}{\partial \dot{\psi}^I}
    = -4i \left( \, (p^+ - hP^- ) \delta^{IJ}
      + \frac{1}{2} h X'^- s^{IJ} \, \right) \psi^J
      + {\cal O} (\psi^3)~.
      \label{pp}
\end{equation}
The detailed form of the Hamiltonian is then
\begin{eqnarray}
H &=& \frac{1}{2p^+} \int^1_0 d \sigma \,
 \bigg\{ \,
     (P^i)^2 + (X'^i)^2 + i \psi^1 \psi'^1 - i \psi^2 \psi'^2
         - \frac{h}{4(p^+)^2} ( P^i +  X'^i )^2 ( P^j - X'^j )^2
                        \nonumber \\
 & & -\frac{ih}{4(p^+)^2} (P^i-X'^i)^2 \psi^1 \psi'^1
     + \frac{i}{8 (p^+)^2} \partial_i h \,
        ( P^k - X'^k)^2 ( P^j + X'^j  ) \psi^1 \gamma_{ij} \psi^1
                        \nonumber \\
 & & + \frac{ih}{4(p^+)^2} (P^i+X'^i)^2 \psi^2 \psi'^2
     + \frac{i}{8 (p^+)^2} \partial_i h \,
        ( P^k + X'^k)^2 ( P^j - X'^j  ) \psi^2 \gamma_{ij} \psi^2 \,
                        \nonumber \\
 & & + \cdots + {\cal O} (\psi^4)
   \, \bigg\} ~,  \label{ham}
\end{eqnarray}
where, in order to make the kinetic term for $\psi^I$ to be of the
canonical form, the following rescaling has been performed.
\begin{equation}
\psi^I  \longrightarrow \frac{1}{2\sqrt{2p^+}} \psi^I ~.
 \label{scale2}
\end{equation}
The dots in (\ref{ham}) denote the terms of order ${\cal O} (h^2)$,
which basically correspond to those of higher derivatives than four.
These are beyond of our interest, since what we are concerned about in
the supergravity side analysis are the terms corresponding to the so
called one-loop exact $F^4$ or four-derivative terms and their
superpartners in the low energy effective action from the SYM side,
which are of linear order in $h$ in Eq.~(\ref{ham}).  Thus, from now
on, we will keep only the terms up to linear order in $h$ in all
expressions in the remaining part of this paper.

The Hamiltonian (\ref{ham}) shows typical interaction terms between
superstring and the background geometry (\ref{back}).  The bosonic
interaction is the spinless one and Eq.~(\ref{har}) tells us that it
has $1/r^6$ behavior.  In SYM side, this type of behavior can be seen
in the perturbative sector of the result of Ref.~\cite{gid121}.
Comparing the result of Ref.~\cite{sch145}, the structure of the
interaction also agrees with that in the SYM side; more precisely, the
Matrix string stress tensor $T^{--}$ in the weak string coupling limit
in that reference.  The term proportional to $\psi^I \gamma^{ij}
\psi^I$ is the spin-orbit interaction term, which has been found and
studied also in other compactifications of DLCQ M theory
\cite{kra199,hyu105}.

Before closing this section, we obtain the equations of motion for
$X^i$ and $\psi^I$ by using the Hamiltonian (\ref{ham}) as the time
evolution operator, which will be used in the next section. By the
way, since we have constraints coming from the momenta of $\psi^I$,
(\ref{pp}), we should take into account them first.  The constraints
are
\begin{equation}
\Phi^I = P^I+\frac{2i}{\sqrt{2 p^+}}
        \left( \, (p^+ - hP^- ) \delta^{IJ}
                   + \frac{1}{2} h X'^- s^{IJ} \,
        \right) \psi^J  ~\approx~ 0 + {\cal O}(\psi^3)~,
\label{const}
\end{equation}
where the rescaling (\ref{scale2}) has been performed. Here $P^-$ and
$X'^-$ should be understood as the solutions of the Virasoro
constraints (\ref{virasoro}).  Then by using the canonical Poisson
brackets
\begin{eqnarray}
\{ X^i (\sigma), P^j (\sigma') \}_{PB}
    &=& \delta^{ij} \delta (\sigma-\sigma')~,
                    \nonumber \\
\{ \psi^{I \alpha}(\sigma), P^{J \beta}(\sigma') \}_{PB}
    &=& -\delta^{IJ} \delta^{\alpha \beta} \delta (\sigma-\sigma')~,
\end{eqnarray}
one can show that $\Phi^I$ are in second class and their time
evolutions give no more constraints.  The usual Dirac procedure for
the constrained system \cite{dir67} then leads to the Dirac bracket,
$\{ \, , \}_D$, consistent with the constraints (\ref{const}).  The
resulting non-vanishing Dirac brackets are,
\begin{eqnarray}
\{ X^i(\sigma), P^j (\sigma') \}_D
 &=& \delta^{ij} \delta(\sigma-\sigma') +{\cal O}(\psi^4)~,
                        \nonumber \\
\{ \psi^{1\alpha}(\sigma), \psi^{1\beta} (\sigma') \}_D
 &=& - i
     \left( 1-\frac{h}{4(p^+)^2} ( P^i - X'^i )^2
     \right) \delta^{\alpha \beta} \delta(\sigma-\sigma')
     +{\cal O}(\psi^2)~,
                        \nonumber \\
\{ \psi^{2\alpha}(\sigma), \psi^{2\beta} (\sigma') \}_D
 &=& - i
     \left( 1 -\frac{h}{4(p^+)^2} ( P^i + X'^i )^2
     \right) \delta^{\alpha \beta} \delta(\sigma-\sigma')
     + {\cal O}(\psi^2)~,
                        \nonumber \\
\{ X^i(\sigma), \psi^{I\alpha}(\sigma') \}_D
 &=& - \frac{h}{4(p^+)^2}
     \left( P^i \delta^{IJ} -  X'^i s^{IJ} \right)
     \psi^{J\alpha} \delta(\sigma-\sigma')+{\cal O} (\psi^3)~,
                        \nonumber \\
\{ P^i(\sigma), \psi^{1\alpha}(\sigma') \}_D
 &=& \frac{1}{8 (p^+)^2} \partial_i h ( P^j - X'^j )^2
       \psi^{1\alpha} \delta(\sigma-\sigma')
                        \nonumber \\
 & &   + \frac{1}{4(p^+)^2}
       \left( \, h ( P^i - X'^i ) \psi^{1\alpha}
       \right) \!(\sigma') \, \partial_\sigma
                           \delta(\sigma-\sigma')
       +{\cal O}(\psi^3)~,
                        \nonumber \\
\{ P^i(\sigma), \psi^{2\alpha}(\sigma') \}_D
 &=& \frac{1}{8 (p^+)^2} \partial_i h ( P^j + X'^j )^2
       \psi^{2\alpha} \delta(\sigma-\sigma')
                        \nonumber \\
 & &   - \frac{1}{4 (p^+)^2}
       \left( \, h ( P^i + X'^i ) \psi^{2\alpha}
       \right) \!(\sigma') \, \partial_\sigma
                \delta(\sigma-\sigma')
       +{\cal O}(\psi^3)~.
       \label{db}
\end{eqnarray}
For the Dirac brackets between $X^i$ and $P^i$, we have omitted the
terms of order ${\cal O}(\psi^2)$ since they are of order ${\cal
O}(h^2)$.  The terms of quadratic order in $\psi$ in the Dirac
brackets between $\psi$'s have not been given due to the fact that, in
order to determine them, we need to know the terms of order ${\cal
O}(\psi^3)$ in the constraints (\ref{const}) which are supposed to
come from the not yet determined ${\cal O}(\psi^4)$ terms in the
Hamiltonian (\ref{ham}).

Having the Dirac brackets, through the time evolutions $\dot{X}^i = \{
X^i, H \}_D$ and $\dot{\psi}^I = \{ \psi, H \}_D$, we now get the
equations of motion for $X^i$ and $\psi^I$ as follows:
\begin{eqnarray}
\dot{X}^i
 &=& \frac{1}{p^+}P^i
     -\frac{h}{2 (p^+)^3} ((P^j)^2+(X'^j)^2) P_i
     +\frac{h}{(p^+)^3} (P^j X'^j) X'^i
                        \nonumber \\
 & & - i \frac{h}{2 (p^+)^3} (P^i - X'^i) \psi^1 \psi'^1
     + i \frac{h}{2 (p^+)^3} (P^i + X'^i) \psi^2 \psi'^2
                        \nonumber \\
 & & + \frac{i}{16 (p^+)^3} \partial_j h
          (P^k - X'^k)^2 \psi^1 \gamma^{ji} \psi^1
     + \frac{i}{8 (p^+)^3} \partial_j h
         (P^i-X'^i) (P^k + X'^k) \psi^1 \gamma^{jk} \psi^1
                        \nonumber \\
 & & + \frac{i}{16 (p^+)^3} \partial_j h
          (P^k + X'^k)^2 \psi^2 \gamma^{ji} \psi^2
     + \frac{i}{8 (p^+)^3} \partial_j h
         (P^i+X'^i) (P^k - X'^k) \psi^2 \gamma^{jk} \psi^2
            \nonumber \\
 & & + {\cal O}(\psi^4)~, \label{xeom}
\end{eqnarray}
\begin{eqnarray}
\dot{\psi}^1
 &=& \frac{1}{p^+} \psi'^1 - \frac{h}{2(p^+)^3} (P^i - X'^i)^2 \psi'^1
    - \frac{1}{8(p^+)^3}\partial_i h (P^j-X'^j)^2(P^i+X'^i)\psi^1
                        \nonumber \\
 & & + \frac{1}{8(p^+)^3}\partial_i h (P^k-X'^k)^2(P^j+X'^j)
    \gamma^{ij} \psi^1 + {\cal O} (\psi^3)~,
            \nonumber \\
\dot{\psi}^2
 &=& -\frac{1}{p^+} \psi'^2 + \frac{h}{2(p^+)^3} (P^i + X'^i)^2 \psi'^2
    - \frac{1}{8(p^+)^3}\partial_i h (P^j+X'^j)^2(P^i-X'^i)\psi^2
                        \nonumber \\
 & & + \frac{1}{8(p^+)^3}\partial_i h (P^k+X'^k)^2(P^j-X'^j)
    \gamma^{ij} \psi^2 + {\cal O} (\psi^3)~. \label{peom}
\end{eqnarray}
We would like to note that these equations of motion in the point
particle limit, i.e., eliminating the terms involving the $\sigma$
derivatives, agrees with those of the eleven dimensional supergraviton
\cite{hyu119} except for the transverse $SO(9)$ invariance rather than
$SO(8)$.

\section{Supersymmetry in Light-Cone Gauge}
\setcounter{equation}{0}
\label{susylc}

The system described by the light-cone gauge Hamiltonian (\ref{ham})
is supersymmetric. In this section, we investigate the supersymmetry
transformation rules for $X^i$ and $\psi^I$, and the supersymmetry
algebra.  To begin with, we consider the supersymmetry preserved by
the background geometry (\ref{back}), which can be seen by looking at
the Killing spinor equation coming from Eq.~(\ref{dpsi}) in the
background, $\delta_\eta \psi^I_\mu=0$:
\[
D_\mu (\omega) \eta^I = 0~.
\]
The solution of this equation shows that $\eta^I$ is of the following
form:
\begin{equation}
\eta^I = f^{-1/4}
  \left( \begin{array}{c} \epsilon^I \\ \epsilon^I \end{array}
  \right) \label{e2ep}
\end{equation}
where $\epsilon^I$ are the 16 component constant spinors. Since
$\eta^1$ ($\eta^2$) has ten dimensional positive (negative)
chirality, $\epsilon^1$ is in the representation ${\bf 8_c}$ of
spin$(8)$ while $\epsilon^2$ is in ${\bf 8_s}$. Thus we see that the
background geometry (\ref{back}) preserves 16 supersymmetries in
total.  As shown in the case of Matrix theory \cite{hyu22}, it is
this abundance of supersymmetry that is responsible for the precise
agreement between the SYM, Matrix string theory, and the supergravity
side calculations in the previous section.

Though they were enough to verify the invariance of superstring
action in Sec.~\ref{susy}, the supersymmetry transformation rules
(\ref{susyt}) do not give us the full supersymmetry structure up to
the quadratic order in terms of $\theta$; we need the transformation
rules expanded up to $\theta^2$ order.  They can be obtained from the
results of \cite{dew209} through the Kaluza-Klein reduction, and, for
the background geometry (\ref{back}), are given by
\begin{eqnarray}
\delta_\eta X^\mu &=& i\bar{\theta}^I \Gamma^\mu \eta^I
        +{\mathcal O}(\theta^3)~,     \nonumber \\
\delta_\eta \theta^I &=& \eta^I
  - \frac{i}{4} (\bar{\theta}^J \Gamma^\mu \eta^J)
      \omega_\mu^{~rs} \Gamma_{rs} \theta^I
  +{\mathcal O} (\theta^4)~.
\end{eqnarray}
In the light-cone gauge specified in Eqs.~(\ref{kfix}) and
(\ref{dfix}), these become
\[
 \delta_\eta X^+ = \delta_\eta X^- = 0 +{\mathcal O}(\theta^3)~,~~~~
 \delta_\eta X^i =i \bar{\theta}^I \Gamma^i \eta^I
        +{\mathcal O}(\theta^3)~,
\]
\begin{equation}
\delta_\eta \theta^I = \eta^I
 - \frac{i}{4} f^{-1} \partial_i h (\bar{\theta}^J \Gamma^- \eta^J)
   \Gamma^- \Gamma^i \theta^I
 + \frac{i}{4} f^{-1} \partial_i h (\bar{\theta}^J \Gamma^i \eta^J)
   \theta^I
 + {\mathcal O} (\theta^4)~,   \label{lcsusy}
\end{equation}
with the supersymmetry parameter $\eta^I$ given by Eq.~(\ref{e2ep})
which satisfies $\Gamma^- \eta^I = 0$.  However, since $\Gamma^+
\eta^I \neq 0$, the above supersymmetry transformation rules break
the $\kappa$-symmetry fixing condition Eq.~(\ref{kfix}), that is,
$\Gamma^+ \delta_\eta \theta^I \neq 0$.  This means that we should
modify the above transformation rules for the correct supersymmetry
in the light-cone gauge.\footnote{This situation has been known in
the study of GS superstring theory in flat background. See, for
example, the chap. 5 of Ref.~\cite{gre87}} In order to preserve the
$\kappa$-symmetry fixing condition, it is natural to use the $\kappa$
transformations for the modification. We may also include the
worldsheet diffeomorphism with parameter $\zeta$ in the modified
supersymmetry transformation rules, for the possibility of breakdown
of the diffeomorphism fixing condition, the light-cone gauge,
Eq.~(\ref{dfix}). Then the modified supersymmetry transformation
$\delta$ is of the following form:
\begin{equation}
\delta = \delta_\eta + \delta_\kappa + \delta_\zeta ~, \label{msusy}
\end{equation}
where $\kappa$ and $\zeta$ are the functions of $\eta$, i.e.,
$\epsilon$, to be determined by the requirement of preserving the
light-cone gauge.

The $\kappa$ transformation rules which also have an expansion in
terms of $\theta$ are again obtained from the eleven dimensional
results of \cite{dew209} through the Kaluza-Klein reduction and, for
the background (\ref{back}), are given by
\begin{eqnarray}
\delta_\kappa X^\mu
 &=& i \bar{\kappa}_+^I \Gamma^\mu \theta^I
     + {\mathcal O} (\theta^3)~,
                    \nonumber \\
\delta_\kappa \theta^I
 &=& \kappa_+^I
    + \frac{i}{4} (\bar{\theta}^J \Gamma^\mu \kappa_+^J)
    \omega_\mu^{~rs} \Gamma_{rs} \theta^I
    + {\mathcal O} (\theta^4)~,
\end{eqnarray}
where $\kappa_+$ is defined in Eq.~(\ref{kp}).  In the light-cone
gauge, Eqs.~(\ref{kfix}) and (\ref{dfix}), we get
\begin{eqnarray}
\delta_\kappa X^+
  &=& -i f^{-1/2} h \,\bar{\kappa}_+^I \Gamma^- \theta^I
        +{\mathcal O}(\theta^3)~,   \nonumber \\
\delta_\kappa X^-
  &=& i f^{-1/2}  \bar{\kappa}_+^I \Gamma^- \theta^I
        +{\mathcal O}(\theta^3)~,     \nonumber \\
\delta_\kappa X^i
  &=& i \bar{\kappa}_+^I \Gamma^i \theta^I
        +{\mathcal O}(\theta^3)~,
                                    \nonumber \\
\delta_\kappa \theta^I
  &=&
  \kappa_+^I
 + \frac{i}{4f} \partial_i h (\bar{\theta}^J \Gamma^- \kappa_+^J)
   \Gamma^- \Gamma^i \theta^I
 - \frac{i}{4f} \partial_i h (\bar{\theta}^J \Gamma^i \kappa_+^J)
   \theta^I
 + {\mathcal O} (\theta^4)~.
\end{eqnarray}
For the superstring case, it is usually convenient to introduce
\begin{equation}
\kappa_m^I \equiv -i \frac{\sqrt{-\gamma}}{2\sqrt{-g}} \Pi_m
                  \cdot \Gamma \kappa^I ~,
                  \label{kmk}
\end{equation}
which allows us to view the transformation parameter $\kappa$ as a
worldsheet vector.  By using Eqs.~(\ref{gam}) and (\ref{proj}), it is
easy to show that $\kappa_m^1$ ($\kappa_m^2$) satisfies the
(anti-)self-dual condition:
\begin{equation}
\kappa^{1 m} = P^{mn,11} \kappa^1_n ~,~~~
 \kappa^{2m} = P^{mn,22} \kappa^2_n ~.
\end{equation}
Therefore, each of these worldsheet vectors has one independent
vector component and hence may be represented as\footnote{ Detailed
expressions of $\kappa^{Im}$ in terms of $\chi^I$ are as follows:
\[
\kappa^{1\tau} = \gamma^{\tau\tau} \chi^1~,~~~
 \kappa^{1\sigma} =\left(\gamma^{\tau\sigma}
       -\frac{1}{\sqrt{-\gamma}}\right)\chi^1~,~~~
\kappa^{2\tau} = \gamma^{\tau\tau} \chi^2~,~~~
 \kappa^{2\sigma} =\left(\gamma^{\tau\sigma}
       +\frac{1}{\sqrt{-\gamma}}\right)\chi^2~.
\]}
\begin{equation}
\kappa^{Im} = 2 P^{m\tau,IJ} \chi^J ~. \label{kmc}
\end{equation}

We now turn to the modified supersymmetry transformation,
(\ref{msusy}), and investigate it order by order in terms of the
anticommuting coordinates.  At the leading order, we first consider
the transformation of $\theta$,
\begin{equation}
\delta^{(0)} \theta^I = \delta^{(0)}_\eta \theta^I +
\delta^{(0)}_\kappa \theta^I + \zeta^{(0)m} \partial_m \theta^I~,
\label{0st}
\end{equation}
where the superscript $(n)$ represents that the explicit order of
$\theta$ (i.e., $\psi$) is $n$.  To preserve the $\kappa$-symmetry
fixing condition, this must satisfy
\begin{equation}
\Gamma^+\delta^{(0)} \theta^I = \Gamma^+ (\delta^{(0)}_\eta \theta^I
+ \delta^{(0)}_\kappa \theta^I) = 0~.   \label{0cc}
\end{equation}
We see that the diffeomorphism parameter $\zeta^{(0)m}$ which is
zeroth order in $\theta$ does not contribute to this consistency
requirement and may be set to zero.  On rewriting Eqs.~(\ref{0st})
and (\ref{0cc}) in terms of $\chi^I$ through Eqs.~(\ref{kmk}) and
(\ref{kmc}), one can show without much difficulty that $\delta^{(0)}
\theta^I$ is given only in terms of $\eta^I$.  If we now express the
resulting transformation $\delta^{(0)} \theta^I$ in terms of the 16
component spinors $\psi^I$ and $\epsilon^I$ by using Eqs.~(\ref{tp})
and (\ref{e2ep}), then we have
\begin{eqnarray}
\delta^{(0)} \psi^1 &=&  f^{-1/4} N^{1i}  \gamma^i \epsilon^1 ~,
                                    \nonumber \\
\delta^{(0)} \psi^2 &=&  f^{-1/4} N^{2i}  \gamma^i \epsilon^2~,
\label{0mpt}
\end{eqnarray}
where we have defined
\begin{equation}
N^{Ii} \equiv
  \frac{\gamma^{\tau\tau} \Pi^i_\tau
         + (\gamma^{\tau\sigma}\mp 1/\sqrt{-\gamma})\Pi^i_\sigma}{
        \gamma^{\tau\tau} \Pi^+_\tau
         + (\gamma^{\tau\sigma}\mp 1/\sqrt{-\gamma})\Pi^+_\sigma}~,
         \label{nii}
\end{equation}
with $\mp$ corresponding to $I=1,2$ respectively. The above
transformation rules are the desired results which preserve the
$\kappa$-symmetry fixing condition. Here we would like to note that,
in the process of calculation, it is crucial to recognize the
following identity satisfied for each $I$:
\begin{equation}
N^{Ii} N^{Ii}=
 -\frac{\gamma^{\tau\tau} \Pi^-_\tau
         + (\gamma^{\tau\sigma}\mp 1/\sqrt{-\gamma})\Pi^-_\sigma}{
        \gamma^{\tau\tau} \Pi^+_\tau
         + (\gamma^{\tau\sigma}\mp 1/\sqrt{-\gamma})\Pi^+_\sigma}~.
         \label{nnii}
\end{equation}
In fact, the identity (\ref{nnii}) is nothing but the covariant
Virasoro constraint as can be verified by a direct calculation.  At
this point, one may worry about the presence of the worldsheet metric
in $N^{Ii}$ about which we have not been concerned so far. However,
as we shall see later, an intriguing fact for $N^{Ii}$ is that,
though the worldsheet metric appears in its definition, the resulting
expression of $N^{Ii}$ is totally independent on the worldsheet
metric.

Let us now consider the leading order modified supersymmetry
transformation of $X^i$, $\delta^{(1)} X^i$.  In this case, there is
a problem related to the light-cone gauge $X^+=2\tau$,
Eq.~(\ref{dfix}).  The transformation $\delta^{(1)} X^+$ is given by
\begin{equation}
\delta^{(1)} X^+ = 4 i f^{-1/2} h \,  ( \psi^I \delta^{(0)} \psi^I)
                  + 2 \zeta^{(1)\tau} ~,
\end{equation}
which does not vanish in general and breaks the light-cone gauge
fixing condition.   In order to recover the light-cone gauge,
$\zeta^{(1)\tau}$ should be chosen such as
\begin{equation}
\zeta^{(1)\tau} = -2 i f^{-1/2} h \,  ( \psi^I \delta^{(0)} \psi^I)~.
\label{d0}
\end{equation}
Then the transformation $\delta^{(1)} X^i$ becomes, up to
diffeomorphism in the $\sigma$ direction,
\begin{equation}
\delta^{(1)} X^i = 4i f^{-1/4} \psi^I \gamma^i \epsilon^I
     - 2 i f^{-1/2} h \, ( \psi^I \delta^{(0)} \psi^I) \dot{X}^i
     + \zeta^{(1)\sigma} X'^i ~.
     \label{0mxt}
\end{equation}
Since the spatial component $\zeta^{(1)\sigma}$ does not appear in
$\delta^{(1)} X^+$ due to $X'^+=0$ in the light-cone gauge, it
remains undetermined.  The requirement of preserving another
condition in the light-cone gauge, (\ref{dfix}), is not helpful as
well for specifying it, since $p^+$ is constant.  As a possible way
of determining it, we consider the closure of supersymmetry algebra,
a property that supersymmetry must satisfy.  If we use the leading
order transformation rules for $\psi^I$ and $X^i$, Eqs.~(\ref{0mpt})
and (\ref{0mxt}), and the equation of motion for $X^i$,
Eq.~(\ref{xeom}), the requirement of $[ \delta_{\epsilon_{(1)}},
\delta_{\epsilon_{(2)}} ] X^i = \xi^m \partial_m X^i$ with $\xi^m$ as
bilinear combinations of $\epsilon_{(1)}$ and $\epsilon_{(2)}$ makes
us to have
\begin{equation}
\zeta^{(1)\sigma} = 2i (p^+)^{-1}
    f^{-1/2} h \, s^{IJ}(\psi^I \delta^{(0)} \psi^J)~.
    \label{d1}
\end{equation}

Through the same procedure performed at the leading order, we can
obtain the next-to-leading order corrections to the modified
supersymmetry transformation.  At this order, there is no need to
consider the corrections to the transformation of $X^i$, since the
next-to-leading order corrections is of order ${\cal O} (\psi^3)$ and
thus beyond of our interest in this paper.  The modified
supersymmetry transformations of $\psi^I$ get non-trivial corrections
at the next-to-leading order and they are obtained as
\begin{eqnarray}
\delta^{(2)} \psi^1
 &=& if^{-5/4} \partial_i h (\psi^I \gamma^i \epsilon^I) \psi^1
 -  2if^{-5/4} (\psi^I \delta^{(0)} \psi^I)
          \partial_i h N^{1j} \gamma^j\gamma^i \psi^1
 + \zeta^{(1)m} \partial_m \psi^1~,~
                                    \nonumber \\
\delta^{(2)} \psi^2
 &=& if^{-5/4} \partial_i h (\psi^I \gamma^i \epsilon^I) \psi^2
 -  2if^{-5/4} (\psi^I \delta^{(0)} \psi^I)
          \partial_i h N^{2j} \gamma^j\gamma^i \psi^2
 + \zeta^{(1)m} \partial_m \psi^2~,
\end{eqnarray}
where $\zeta^{(1)m}$ are given by Eqs.~(\ref{d0}) and (\ref{d1}).

Up to the quadratic order in $\psi$, we have obtained all the
informations for the supersymmetry transformations in the light-cone gauge.  By gathering the order by order results, the full
modified supersymmetry transformation rules preserving the light-cone
gauge and $\kappa$-symmetry fixing conditions are given by
\begin{eqnarray}
\delta X^i &=& \delta^{(1)} X^i + {\cal O} (\psi^3)~,
                            \nonumber \\
\delta \psi^I &=& \delta^{(0)} \psi^I + \delta^{(2)} \psi^I
        +{\cal O} (\psi^4)~.
\end{eqnarray}
The detailed form of the above transformations are obtained by
doing the rescalings for $\psi^I$, Eqs.~(\ref{scale1}) and (\ref{scale2}), using the equations of motion for $X^i$ and $\psi^I$,
Eqs.~(\ref{xeom}) and (\ref{peom}), and the following expansions
for $N^{Ii}$, Eq.~(\ref{nii}):
\begin{eqnarray}
f^{-1/2} N^{1i}
 &=& \frac{1}{2p^+} (P^i+X'^i)
     - \frac{1}{8(p^+)^3} (P^j-X'^j)^2(P^i+X'^i)
                        \nonumber \\
 & & + \frac{i}{4(p^+)^3} h (P^i+X'^i) (\psi^2 \psi'^2)
     + \frac{i}{8(p^+)^3} \partial_j h P^k X'^k
        (\psi^1 \gamma^{ij}\psi^1)
                        \nonumber \\
 & & + \frac{i}{16(p^+)^3} \partial_j h (P^i+X'^i)
    (P^k \delta^{IJ} + X'^k s^{IJ})(\psi^I \gamma^{jk} \psi^J)
     + {\cal O} (\psi^4) ~,
                \nonumber \\
f^{-1/2} N^{2i}
 &=& \frac{1}{2p^+} (P^i-X'^i)
     - \frac{1}{8(p^+)^3} (P^j+X'^j)^2(P^i-X'^i)
                        \nonumber \\
 & & - \frac{i}{4(p^+)^3} h (P^i-X'^i) (\psi^1 \psi'^1)
     - \frac{i}{8(p^+)^3} \partial_j h P^k X'^k
        (\psi^2 \gamma^{ij}\psi^2)
                        \nonumber \\
 & & + \frac{i}{16(p^+)^3} \partial_j h (P^i-X'^i)
    (P^k \delta^{IJ} + X'^k s^{IJ})(\psi^I \gamma^{jk} \psi^J)
     + {\cal O} (\psi^4) ~.
\end{eqnarray}
As alluded before, we see that $N^{Ii}$ do not have the dependence
on the worldsheet metic.  The final expressions we have got are
\begin{eqnarray}
\delta X^i
 &=&  \frac{2i}{\sqrt{2p^+}} \psi^I \gamma^i \epsilon^I
    - \frac{i}{ 2\sqrt{2p^+} (p^+)^2 } h (P^i-X'^i)(P^j+X'^j)
     ( \psi^1 \gamma^j  \epsilon^1)
                        \nonumber \\
 & & - \frac{i}{ 2\sqrt{2p^+} (p^+)^2 } h (P^i+X'^i)(P^j-X'^j)
    ( \psi^2 \gamma^j  \epsilon^2) + {\mathcal O} (\psi^3)~,
                \nonumber \\
\delta \psi^1
 &=& \frac{2}{\sqrt{2p^+}} (P^i+X'^i) \gamma^i \epsilon^1
    -\frac{1}{2\sqrt{2p^+}(p^+)^2} h (P^j-X'^j)^2 (P^i+X'^i)
        \gamma^i\epsilon^1
                        \nonumber \\
 & &+ \frac{i}{\sqrt{2p^+}(p^+)^2} h (P^i+X'^i) (\psi^2 \psi'^2)
        \gamma^i \epsilon^1
    -\frac{i}{\sqrt{2p^+}(p^+)^2} h (P^i-X'^i)
        (\psi^2 \gamma^i \epsilon^2) \psi'^1
                        \nonumber \\
 & & - \frac{i}{2\sqrt{2p^+}(p^+)^2} \partial_i h (P^k X'^k)
        (\psi^1 \gamma^{ij} \psi^1) \gamma^j \epsilon^1
                        \nonumber \\
 & & + \frac{i}{4\sqrt{2p^+}(p^+)^2} \partial_i h (P^j+X'^j)
        (P^k \delta^{IJ}+X'^k s^{IJ})
        (\psi^I \gamma^{ik} \psi^J) \gamma^j \epsilon^1
                        \nonumber \\
 & & -\frac{i}{4\sqrt{2p^+}(p^+)^2} \partial_i h (P^j+X'^j)
        (P^k\delta^{IJ} +X'^k s^{IJ})(\psi^I \gamma^k \epsilon^J)
        \gamma^j \gamma^i \psi^1
     + {\mathcal O} (\psi^4)~,
                \nonumber \\
\delta \psi^2
 &=& \frac{2}{\sqrt{2p^+}} (P^i-X'^i) \gamma^i \epsilon^2
    -\frac{1}{2\sqrt{2p^+}(p^+)^2} h (P^j+X'^j)^2 (P^i-X'^i)
        \gamma^i\epsilon^2
                        \nonumber \\
 & &- \frac{i}{\sqrt{2p^+}(p^+)^2} h (P^i-X'^i) (\psi^1 \psi'^1)
        \gamma^i \epsilon^2
    +\frac{i}{\sqrt{2p^+}(p^+)^2} h (P^i+X'^i)
        (\psi^1 \gamma^i \epsilon^1) \psi'^2
                        \nonumber \\
 & & + \frac{i}{2\sqrt{2p^+}(p^+)^2} \partial_i h (P^k X'^k)
        (\psi^2 \gamma^{ij} \psi^2) \gamma^j \epsilon^2
                        \nonumber \\
 & & + \frac{i}{4\sqrt{2p^+}(p^+)^2} \partial_i h (P^j-X'^j)
        (P^k \delta^{IJ}+X'^k s^{IJ})
        (\psi^I \gamma^{ik} \psi^J) \gamma^j \epsilon^2
                        \nonumber \\
 & & -\frac{i}{4\sqrt{2p^+}(p^+)^2} \partial_i h (P^j-X'^j)
        (P^k\delta^{IJ} +X'^k s^{IJ})(\psi^I \gamma^k \epsilon^J)
        \gamma^j \gamma^i \psi^2
     + {\mathcal O} (\psi^4)~.
 \label{mst}
\end{eqnarray}
Obviously, these supersymmetry transformations are global from the
worldsheet point of view, since the transformation parameters
$\epsilon^I$ are constants, while the starting supersymmetry
transformations (\ref{lcsusy}) are local in target space-time.  If we
set $h=0$ for a moment, the transformations Eq.~(\ref{mst}) are
nothing but those of scalar multiplets of two dimensional ${\cal
N}=(8,8)$ SYM theory obtained in early days of GS superstring theory
\cite{gre444}. The interpretation of this is clear in the context of
DLCQ M theory, though it is not so in the original GS superstring
theory itself.  The transformation rules for the case of $h=0$ are
those of Matrix string theory at the tree level corresponding to a
free superstring. The $h$ dependent terms in Eq.~(\ref{mst}) are
due to the one-loop corrections to the Matrix string theory for the
two superstring background.

With the modified supersymmetry transformation rules (\ref{mst}), it
is a straightforward task to investigate the supersymmetry algebra.
Since we have not determined the terms of the order ${\cal O}
(\psi^3)$ in $\delta X^i$, it is not possible to check the
supersymmetry algebra for $X^i$ to quadratic order in terms of
$\psi$.  On the contrary, $\delta \psi^I$ leads to the non-trivial
check.  By using the $SO(8)$ Fierz identity for the spinors with the
same $SO(1,9)$ chiralities,
\begin{equation}
(\psi \gamma^i \epsilon_{(1)}) \gamma^i \epsilon_{(2)}
=  (\epsilon_{(1)} \epsilon_{(2)})\psi
 -\frac{1}{4}(\epsilon_{(1)} \gamma^{ij} \epsilon_{(2)})
        \gamma^{ij} \psi ~,
\end{equation}
we can show that, up to the equations of motion,
\begin{equation}
[ \delta_{\epsilon_{(1)}},\delta_{\epsilon_{(2)}} ] \psi^I
= \xi^m \partial_m \psi^I +{\cal O} (\psi^3)~,
\label{qq}
\end{equation}
where the worldsheet translation parameters $\xi^m$ are given by
\begin{equation}
\xi^\tau = 4i ( \epsilon^1_{(1)} \epsilon^1_{(2)}
               +\epsilon^2_{(1)} \epsilon^2_{(2)} )~,~~~
\xi^\sigma = 4i (p^+)^{-1}
        ( \epsilon^1_{(1)} \epsilon^1_{(2)}
             -\epsilon^2_{(1)} \epsilon^2_{(2)} )~.
\end{equation}
The algebra (\ref{qq}) is what we want in the two dimensional theory
and corresponds to the anticommutation relation between the would-be
supercharges $Q^I$ generating the transformations (\ref{mst}); $\{
Q^I, Q^J \} \propto \delta^{IJ} H + s^{IJ} P$ where $H$ and $P$ are
the translation generators in two dimensions.  Besides of the algebra
(\ref{qq}), another thing to be investigated is the supersymmetry
transformation property of the system described by the Hamiltonian
(\ref{ham}), which is obtained as
\begin{equation}
\delta_{\epsilon^I} H = 0 + {\cal O} (\psi^3) \label{qh}~.
\end{equation}
As usual, this means that the system is supersymmetric. In other
words, the supercharges are conserved: $[ Q^I, H ] = 0$.

\section{Discussion}
\setcounter{equation}{0}
\label{dis}

We have studied the light-cone superstring dynamics on the
gravitational wave background corresponding to the Matrix string
theory and investigated the structure of supersymmetry.  This is the
supergravity side analysis of the Matrix string theory.  Basically due
to the enough amount of supersymmetry preserved by the background, 16
supersymmetries, the results on dynamics have agreed with those
obtained from the Matrix string theory, the SYM side.  The
supersymmetry transformation rules in the light-cone gauge have been
identified with those of the low energy one-loop effective action of
Matrix string theory for two superstring background in weak string
coupling.  The importance of our results is that the supersymmetry
transformation rules obtained in this paper may provide an alternative
approach to determine some parts of the higher loop-corrections to the
low energy effective action of Matrix string theory without explicit
loop calculations.  The full expansion of transformation rules up to
16th order in terms of the anticommuting coordinates $\psi^I$ will
give us more information about the dynamics of Matrix string theory or
the light-cone superstring on the gravitational wave background.

In our formulation of light-cone superstring, we have fixed two
worldsheet diffeomorphisms by choosing two phase space variables $X^+$
and $P^+$, the light-cone gauge (\ref{dfix}).  An intriguing fact is
that the worldsheet metric has not appeared in the various final
results and hence we have not needed to worry about how it is fixed
according to the light-cone gauge.  If it played a role in any way,
our formulation would be quite complicated.  It is hard to believe
that this situation is an accident.  We expect that the independence
on the worldsheet metric holds also in other supergravity backgrounds,
at least as far as the same kind of calculations done in this paper
is concerned.

For supersymmetry in the light-cone gauge, we have not tried to get
the conserved supercharges $Q^I$ after obtaining the supersymmetry
transformation rules (\ref{mst}).  This is because $Q^I$ generating
(\ref{mst}) should have an expansion up to the order ${\cal O}
(\psi^3)$, which requires the knowledge about the terms of the order
${\cal O}(\psi^4)$ in the light-cone Hamiltonian (\ref{ham}). (Recall
that the terms of that order in the light-cone Hamiltonian have not
yet been determined.)  Furthermore, since the Dirac brackets
(\ref{db}) are used to obtain the supersymmetry algebra and the
transformation rules, i.e., $\delta = i \epsilon^I \{ Q^I, ~\}_D$, we
should know the terms of the order ${\cal O} (\psi^3)$ in the
constraints (\ref{const}) to get the Dirac brackets with the expansion
up to the required order.  For the case of the Matrix theory, one of
the present authors \cite{hyu119} has pointed out that the
supercharges do not receive corrections of qubic or possibly higher
order in anticommuting coordinates and have the same form with that
for the flat case, while the Dirac brackets get corrections.  If we
believe that this is also the case in the present study, the
supercharges $Q^I$ are simply given by
\[
Q^I = \sqrt{2} \, (p^+)^{-1/2} \int d\sigma \,
      (P^i \delta^{IJ} + X'^i s^{IJ} ) \gamma^i \psi^J~.
\]
Indeed, if the Dirac brackets (\ref{db}) are used, these supercharges
generate the leading order terms of the light-cone supersymmetry
transformation rules Eqs.~(\ref{0mpt}) and (\ref{0mxt}), i.e.,
$\delta^{(0)} \psi^I$ and $\delta^{(1)} X^i$.  It is expected that the
full transformation rules of Eq.~(\ref{mst}) would be generated if the
desired corrections were included in the Dirac brackets (\ref{db}).

The final point we would like to discuss is that the formulation given
in this paper is essentially perturbative in view of the Matrix string
theory.  Off the conformal point, the Matrix string theory has the
electric sector which describes the dynamics of D0-branes.  In
Ref.~\cite{gid121}, the process of exchanging D0-branes between two
superstrings in the transverse direction has been considered.  It is
basically the instanton-like process, and is thus non-perturbative. In
the supergravity side, how to see this process and more generally how
to extend the present formulation to the off conformal regime of
Matrix string theory remains an interesting problem.

%
%

\section*{Acknowledgments}

We would like to thank Youngjai Kiem, Sangmin Lee, Jeong-Hyuck Park
and Won Tae Kim for useful discussions and comments. One of us (S.H.)
was supported in part by grant No. 2000-1-11200-001-3 from the Basic
Research Program of the Korea Science and Engineering Foundation.


\end{document}